\documentclass{article}
\usepackage{ltwol2e}

\def\Journal#1#2#3#4{{#1} {\bf #2}, #3 (#4)}

\def\NPB{{\em Nucl. Phys.} B}
\def\PLB{{\em Phys. Lett.}  B}

\def\PRD{{\em Phys. Rev.} D}

\def\be{\begin{equation}}
\def\ee{\end{equation}}
\def\bea{\begin{eqnarray}}
\def\eea{\end{eqnarray}}
\begin{document}

\title{Vilkovisky-DeWitt Effective Potential and the Higgs-Mass Bound}

\author{Guey-Lin Lin and Tzuu-Kang Chyi}

\address{Institute of Physics, National Chiao-Tung University,
Hsinchu, Taiwan, R.O.C.\\E-mail: glin@cc.nctu.edu.tw}   

%%%%%%%%%%%%%%%%%%%%%%%%%%%%%%%%%%%%%%%%%%%%%%%%%%%%%%%%%%%%%%
% You may repeat \author \address as often as necessary      %
%%%%%%%%%%%%%%%%%%%%%%%%%%%%%%%%%%%%%%%%%%%%%%%%%%%%%%%%%%%%%%

\twocolumn[
\maketitle\abstracts{ We compute the Vilkovisky-DeWitt effective 
potential of a simplified 
version of the Standard Electroweak Model, where all charged boson fields
as well as the bottom-quark field are neglected. The effective potential
obtained in this formalism is gauge-independent. 
We derive from the effective potential the 
mass bound of the Higgs boson. The result is compared 
to its counterpart obtained from the ordinary
effective potential.   }
]

\section{Introduction}

The gauge dependence of the effective potential was first pointed out 
by Jackiw in early seventies\cite{jackiw}. This finding  
raised concerns on the physical
significance of the effective potential. In a later work by
Dolan and Jackiw \cite{DJ}, the effective potential of scalar
QED was calculated in a set of $R_{\xi}$ gauges. It was concluded that only 
the limiting unitary gauge gives sensible result on
spontaneous symmetry-breaking. 
This difficulty was partially resolved 
by the work of Nielsen\cite{niel}. In his paper, Nielsen derived 
a simple identity characterizing  
the mean-field and the gauge-fixing-parameter dependences of the 
effective potential, namely,
\begin{equation}
(\xi{\partial \over \partial \xi}+C(\phi,\xi)
{\partial\over \partial \phi})V(\phi, \xi)=0,
\end{equation}    
where $\xi$ is the parameter appearing in the gauge-fixing term 
$L_{gf}=-{1\over 2 \xi}
(\partial_{\mu}A^{\mu})^2$. 
The above identity implies that 
the local extrema of $V$ for different $\xi$ are located along 
the same characteristic
curve on $(\phi,\xi)$ plane, which  satisfies 
$d\xi={d\phi\over C(\phi,\xi)/\xi}$. 
Hence covariant gauges with different $\xi$ are equally
good for computing $V$. On the other hand, a choice of the multi-parameter
gauge\cite{DJ} $L_{gf}=-{1\over 2 \xi}
(\partial_{\mu}A^{\mu}+\sigma \phi_1 +\rho \phi_2)^2$ would break the 
homogeneity of Eq. (1)\cite{niel}. 
Hence effective potential calulated in this
gauge has no physical significance. 

Recently it was pointed out\cite{LW} that the Higgs mass bound as derived
from the effective potential is gauge-dependent. 
The gauge dependence enters in the 
calculation of one-loop effective potential, a quantity that is crucial for
the determination of the Higgs mass bound.
Boyanovsky, Loinaz and Willey has proposed a resolution\cite{BLW} to the 
gauge dependence of the Higgs mass bound. Their approach is based upon
the {\it Physical
Effective Potential} constructed as the expectation value of 
the Hamiltonian in physical states\cite{BBHL}. The effective potential of
the abelian Higgs model is computed explicitly as an illustration.
However, this formalism requires the identification
of first-class constraints in the theory and a projection to 
the physical states. Such a procedure necessarily breaks the manifest
Lorentz invariance of the theory. Consequently we expect it is 
highly non-trivial to apply this formalism to the Standard Model(SM).

In our work\cite{lc}, we introduce the formalism of Vilkovisky and DeWitt
\cite{vil,dw2} for constructing an gauge-independent effective potential,
and therefore obtaining a gauge-independent lower bound for the Higgs mass.
We present the idea with a toy model\cite{LSY} 
which corresponds to neglect all charged boson fields in the SM.
The generalization to the full SM is straightforward. 
In fact,  
the applicability of Vilkovisky-DeWitt formulation 
to non-abelian gauge theories has been
extensively demonstrated in literatures\cite{rebhan}.         

The outline of this presentation is as follows. In Section 2, we briefly
review the formalism of Vilkovisky and DeWitt using scalar QED as 
an example. We shall illustrate that the effective action of Vilkovisky
and DeWitt is equivalent to the ordinary effective action
constructed in the Landau-DeWitt gauge\cite{FT}. In Section 3, we calculate 
the effective potential of the simplified standard model, and the
relevant renormalization constants of the theory using the Landau-DeWitt
gauge. 
The effective potential
is then extended to large vacuum expectation value of the scalar field
by means of renormalization group analyses. In Section 4, the mass bound
of the Higgs boson is derived and compared to that given by oridinary
effective action. Section 5 is the conclusion.

\section{Vilkovisky-DeWitt Effective Action of Scalar QED}

The formulation of Vilkovisky-DeWitt efective action is motivated
by the parametrization dependence of the ordinary effective action,
which can be written generically as
\begin{eqnarray}
\exp{i\over \hbar}\Gamma[\Phi]&=&\exp{i\over \hbar}
(W[j]+\Phi^i{\delta \Gamma
\over \delta \Phi^i})\nonumber \\
&=& \int [D\phi]\exp{i\over \hbar}(S[\phi]+(\Phi^i-\phi^i)\cdot
{\delta \Gamma
\over \delta \Phi^i}).
\label{INTEG}
\end{eqnarray}
The parametrization dependence of the ordinary effective action  
arises because the difference $\eta^i\equiv (\Phi^i-\phi^i)$ is not a 
vector
in the field configuration space, hence the product
$\eta^i \cdot {\delta \Gamma
\over \delta \Phi^i}$ is not a scalar under reparametrization. 
The remedy to this problem is to replace $-\eta^i$ 
with a two-point function $\sigma^i(\Phi,\phi)$ \cite{vil,dw2,com1} 
which, at the
point $\Phi$, 
is tangent to the geodesic connecting $\Phi$ and $\phi$.
The precise form of $\sigma^i(\Phi,\phi)$ depends on the 
connection $\Gamma^i_{jk}$ of the configuration space. It is easy to 
show that\cite{kun} 
\begin{equation}
\sigma^i(\Phi, \phi)=-\eta^i-{1\over 2}\Gamma^i_{jk}\eta^j \eta^k
+ O(\eta^3).
\end{equation}
For scalar QED described
by the Lagrangian:
\begin{eqnarray}
L&=&-{1\over 4}F_{\mu\nu}F^{\mu\nu}+(D_{\mu}\phi)^{\dagger}
(D^{\mu}\phi)\nonumber \\
&-& \lambda (\phi^{\dagger}\phi-\mu^2)^2,
\label{SQED}
\end{eqnarray}
with $D_{\mu}=\partial_{\mu}+ie A_{\mu}$ and $\phi={\phi_1+i\phi_2\over 
\sqrt{2}}$, we have 
\begin{equation}
\Gamma^i_{jk}= {i\brace j k}+T^i_{jk},
\end{equation}
where ${i\brace j k}$ is the Christoffel symbol of the field 
configuration space which has the following
metric:
\begin{eqnarray}
G_{\phi_a(x)\phi_b(y)}&=&\delta_{ab}\delta^4(x-y),\nonumber \\ 
G_{A_{\mu}(x)A_{\nu}(y)}&=&-g^{\mu\nu}\delta^4(x-y),\nonumber \\
G_{A_{\mu}(x)\phi_a(y)}&=&0.
\end{eqnarray}
We note that the metric of the field configuration space is determined 
by the quadratic part of the Lagrangian according to the prescription
of Vilkovisky\cite{vil}. In the above flat-metric, we have ${i\brace jk}=0$. 
However, the
Christtoffel symbols would be non-vanishing in the parametrization with
polar variables $\rho$ and $\chi$ such that
$\phi_1=\rho \cos\chi$ and $\phi_2=\rho \sin\chi$.  
$T^i_{jk}$ is a quantity pertinent to generators $g^i_{\alpha}$ 
of the gauge transformation. Explicitly, we have\cite{vil,kun}
\begin{equation}
T^i_{jk}=-B^{\alpha}_jD_kg^i_{\alpha}+{1\over 2}
g^{\rho}_{\alpha}D_{\rho}K^i_{\beta}
B^{\alpha}_jB^{\beta}_k+ j\leftrightarrow k,
\end{equation}  
where $B^{\alpha}_k=N^{\alpha\beta}g_{k\beta}$ with
$N^{\alpha\beta}$ being the inverse of $N_{\alpha\beta}\equiv 
g^k_{\alpha}g^l_{\beta}G_{kl}$. In scalar QED, the generators
$g^i_{\alpha}$ are given by
\begin{eqnarray}
g^{\phi_a(x)}_y&=&-\epsilon^{ab}\phi_b(x)\delta^4(x-y),\nonumber \\
g^{A_{\mu}(x)}_y&=&-\partial_{\mu}\delta^4(x-y).
\end{eqnarray}
with $\epsilon^{12}=1$.
The one-loop effective action of scalar QED can be calculated from Eq.
(\ref{INTEG}) with the quantum fluctuations $\eta^i$ replaced by
$\sigma^i(\Phi, \phi)$. The result is written as\cite{kun}:
\begin{equation}
\Gamma[\Phi]=S[\Phi]-{i\hbar\over 2}\ln\det G+
{i\hbar\over 2}\ln\det \tilde{D}^{-1}
_{ij},
\label{ACTION}
\end{equation}
where $S[\Phi]$ is the tree-level effective action; $\ln\det G$ arises 
from the function space measure $[D\phi]\equiv \prod_x d\phi(x)
\sqrt{\det G}$, and $\tilde{D}^{-1}_{ij}$ is the modified inverse-propagator:
\begin{equation}
\tilde{D}^{-1}_{ij}={\delta^2 S\over \delta\Phi^i \delta\Phi^j}
-\Gamma^k_{ij}[\Phi]{\delta S\over \delta \Phi^k}.
\end{equation}
To study the symmetry-breaking behaviour of the theory, we focus on
the effective potential which can be obtained from 
$\Gamma[\Phi]$ by setting the classical fields $\Phi$'s to constants. 

The Vilkovisky
-DeWitt effective potential of scalar QED has been calculated 
in various gauges and different parametrizations of scalar fields
\cite{FT,kun,rt}.     
The results all agree with one another. In this work, we 
calculate the effective potential and other relevant
quantities in Landau-DeWitt gauge\cite{com2}, which is characterized by the
gauge-fixing term:
%\begin{equation}
$L_{gf}=-{1\over 2\xi}(\partial_{\mu}B^{\mu}-ie\eta^{\dagger}
\Phi+ie\Phi^{\dagger}\eta)^2$,
%\end{equation}
with $\xi\to 0$. In $L_{gf}$,
$B^{\mu}\equiv A^{\mu}-A^{\mu}_{cl}$, and $\eta \equiv \phi-\Phi$
are quantum fluctuations while $A^{\mu}_{cl}$ and $\Phi$ are classical 
fields. 
The advantage of performing calculations in the Landau-DeWitt gauge
is that $T^i_{jk}$ vanishes\cite{FT} in this case.
In other words, Vilkovisky-DeWitt formalism coincides with the 
conventional one in the Landau-DeWitt gauge.
 
For computing the effective potential, we choose
$A^{\mu}_{cl}=0$ and $\Phi={\rho_{cl}\over \sqrt{2}}$, i.e. the imaginary
part of $\Phi$ is set to zero. In this set of background fields, 
$L_{gf}$ can be written as
\begin{equation}
L_{gf}=-{1\over 2\xi}\left(\partial_{\mu}B^{\mu}\partial_{\nu}B^{\nu}
-2e\rho_{cl}\chi\partial_{\mu}B^{\mu}+e^2\rho_{cl}^2\chi^2\right),
\label{GAUGE}
\end{equation}
where $\chi$ is the quantum field defined by $\eta={\rho+i\chi\over 
\sqrt 2}$. We note that $B_{\mu}-\chi$ mixing in $L_{gf}$ is 
$\xi$ dependent, and therefore would not cancell out 
the corresponding mixing term in the classical Lagrangian of 
Eq. (\ref{SQED}). This induces the mixed-propagator such as
$<0\vert T(A_{\mu}(x)\chi(y)) \vert 0>$ 
or $<0\vert T(\chi(x)A_{\mu}(y)) \vert 0>$. The Faddeev-Popov ghost
Lagrangian is given by
\begin{equation}
L_{FP}=\omega^*(-\partial^2-e^2\rho_{cl}^2)\omega.
\label{FADPOP}
\end{equation}
With each part of the Lagrangian determined, we are ready to
compute the effective potential. Since we choose a flat-metric,
the one-loop effective potential is completely determined by 
the modified inverse propagators $\tilde{D}^{-1}_{ij}$\cite{grassmann}. 
From 
Eqs. (\ref{SQED}), (\ref{GAUGE}) and (\ref{FADPOP}), we arrive at
\begin{eqnarray}
\tilde{D}^{-1}_{B_{\mu}B_{\nu}}&=&(-k^2+e^2\rho_0^2)g^{\mu\nu}
+(1-{1\over \xi})k^{\mu}k^{\nu},\nonumber \\
\tilde{D}^{-1}_{B_{\mu}\chi}&=&ik^{\mu}e\rho_0(1-{1\over \xi}),
\nonumber \\
\tilde{D}^{-1}_{\chi\chi}&=&(k^2-m_G^2-{1\over \xi}e^2\rho_0^2),
\nonumber \\
\tilde{D}^{-1}_{\rho\rho}&=&(k^2-m_H^2),\nonumber \\
\tilde{D}_{\omega^*\omega}&=&(k^2-e^2\rho_0^2)^{-2},
\label{PROP}
\end{eqnarray} 
where we have set $\rho_{cl}=\rho_0$, which is a space-time independent 
constant, and defined 
$m_G^2= \lambda (\rho_0^2-2\mu^2)$,
$m_H^2=\lambda (3\rho_0^2-2\mu^2)$. 
Using the definition $\Gamma[\rho_0]=(2\pi)^4\delta^4(0)V_{eff}(\rho_0)$
along with Eqs. (\ref{ACTION}) and
(\ref{PROP}), and taking the limit $\xi\to 0$, we obtain
$V_{eff}(\rho_0)=V_{tree}(\rho_0)+V_{1-loop}(\rho_0)$ with
\begin{eqnarray}
V_{1-loop}(\rho_0)&=&{-i\hbar\over 2}\int {d^nk\over (2\pi)^n}
\ln[(k^2-e^2\rho_0^2)^{n-3}\nonumber \\
&\times&(k^2-m_H^2)(k^2-m_+^2)(k^2-m_-^2)],
\label{EFFECTIVE}
\end{eqnarray}
where $m_+^2$ and $m_-^2$ are solutions of the quadratic equation
$(k^2)^2-(2e^2\rho_0^2+m_G^2)k^2+e^4\rho_0^4=0$. In the above equation, the
gauge-boson's degree of freedom has been continued to $n-3$ in order to
preserve the relevant Ward identities. Our expression of $V_{1-loop}(\rho_0)$
agree with previous results obtained in the unitary gauge\cite{rt}. 
One could also calculate the effective potential 
in the {\it ghost-free} covariant 
gauges with $L_{gf}=-{1\over 2\xi}(\partial_{\mu}B^{\mu})^2$. 
The cancellation of gauge-parameter dependence in the effective 
potential has been demonstrated in 
the case of massless
scalar-QED with $\mu^2=0$\cite{FT,kun}. It can be easily extended to the 
massive case. 

It is instructive to rewrite Eq. (\ref{EFFECTIVE}) as
\begin{eqnarray}
V_{1-loop}(\rho_0)&=&{\hbar \over 2}\int {d^{n-1}\vec{k}\over (2\pi)^{n-1}} 
\left(
(n-3)\omega_B(\vec{k})+\omega_H(\vec{k})\right. \nonumber \\
&+&\left.\omega_+(\vec{k})
+\omega_-(\vec{k})
\right),
\end{eqnarray}
where $\omega_B(\vec{k})=\sqrt{\vec{k}^2+e^2\rho_0^2}$, 
$\omega_H(\vec{k})=\sqrt{\vec{k}^2+m_H^2}$ and $\omega_{\pm}(\vec{k})
=\sqrt{\vec{k}^2+m_{\pm}^2}$. One can see that $V_{1-loop}$ is 
a sum of the zero-point energies of four excitations with masses
$m_B\equiv e\rho_0$, $m_H$, $m_+$ and $m_-$. Since there 
are precisely four physical degrees of freedom in scalar QED,
we see that Vilkovisky-DeWitt effective potential does exhibit a 
correct number of physical degrees of freedom.

\section{Vilkovisky-DeWitt Effective Potential of the Simplified Standard
Model}

In this section, we compute the effective potential of the simplified 
standard model where charged boson fields and all fermion fields except 
the top quark field are discarded. The gauge interactions of this model are
prescribed by the following covariant derivatives\cite{LSY}:
\begin{eqnarray}
D_{\mu}t_L&=&(\partial_{\mu}+ig_LZ_{\mu}-{2\over 3}ieA_{\mu})t_L,
\nonumber \\
D_{\mu}t_R&=&(\partial_{\mu}+ig_RZ_{\mu}-{2\over 3}ieA_{\mu})t_R,
\nonumber \\
D_{\mu}\phi&=&(\partial_{\mu}+i(g_L-g_R)Z_{\mu})\phi,
\end{eqnarray}
where $g_L=(-g_1/2+g_2/3)$, $g_R=g_2/3$ with $g_1=g/\cos\theta_W$ and
$g_2=2e\tan\theta_W$. Clearly this toy model exhibits a $U(1)_A
\times U(1)_Z$
symmetry where each $U(1)$ symmetry is 
associated with a neutral gauge boson. The $U(1)_Z$-charges of $t_L$, $t_R$
and $\phi$ are related in such a way that the following 
Yukawa interactions are
invariant under $U(1)_A\times U(1)_Z$:
\begin{equation}
L_Y=-y\bar{t}_L\phi t_R-y\bar{t}_R\phi^* t_L.
\end{equation}
Since Vilkosvisky-DeWitt effective action coincides with ordinary effective 
action calculated in the Landau-DeWitt gauge, we hence
calculate the effective potential in this gauge which has
\begin{eqnarray}
L_{gf}=&-&{1\over 2\alpha}(\partial_{\mu}Z^{\mu}+{ig_1\over 2}\eta^{\dagger}
\Phi-{ig_1\over 2}\Phi^{\dagger}\eta)^2\nonumber \\
&-&{1\over 2\beta}(\partial_{\mu}A^{\mu})^2,
\label{GF}
\end{eqnarray}
with $\alpha, \ \beta\to 0$. 
We note that $A^{\mu}$ and $Z^{\mu}$ are quantum fluctuations associated 
with the photon and the $Z$ boson. Their classical backgrounds can be 
set to zero
for computing the effective potential. Following the method 
of the previous section, we obtain
\begin{eqnarray}
V_{VD}(\rho_0)&=&{\hbar \over 2}\int {d^{n-1}\vec{k}\over (2\pi)^{n-1}} 
\left((n-3)\omega_Z(\vec{k})+\omega_H(\vec{k})\right.\nonumber \\
&+&\left. \omega_+(\vec{k})+\omega_-(\vec{k})
-4\omega_F(\vec{k})\right),
\label{POTENTIAL2}
\end{eqnarray}
with $\omega_i(\vec{k})=\sqrt{\vec{k}^2+m_i^2}$ where
$m_Z^2={g_1^2\over 4}\rho_0^2$, $m_{\pm}^2=m_Z^2+{1\over 2}(m_G^2\pm 
m_G\sqrt{m_G^2+4m_Z^2})$ and $m_F^2\equiv m_t^2={y^2\rho_0^2\over 2}$.
Performing the integration in Eq. (\ref{POTENTIAL2})
and subtracting the infinities with $\overline{MS}$ prescription, we obtain
\begin{eqnarray}
V_{VD}(\rho_0)&=&{1\over 64\pi^2}(m_H^4\ln{m_H^2\over \kappa^2} 
+m_Z^4\ln{m_Z^2\over \kappa^2}\nonumber \\
&+& m_+^4\ln{m_+^2\over \kappa^2}+
m_-^4\ln{m_-^2\over \kappa^2}-4m_t^4\ln{m_t^2\over \kappa^2})\nonumber \\
&-&{1\over 128\pi^2}(3m_H^4+5m_Z^4+3m_G^4 \nonumber \\
&+& 12m_G^2m_Z^2
-12m_t^4).
\end{eqnarray}
Although $V_{VD}(\rho_0)$ is obtained in the Landau-DeWitt gauge, we
should stress that any other gauge with non-vanishing $T^i_{jk}$ should lead to
the same result. For later comparisons, let us write down the ordinary 
effective
potential in the {\it ghost-free} Landau gauge\cite{DJ}
(equivalent to removing the scalar part of Eq. (\ref{GF})):
\begin{eqnarray}
V_{L}(\rho_0)&=&{\hbar \over 2}\int {d^{n-1}\vec{k}\over (2\pi)^{n-1}} 
\left((n-1)\omega_Z(\vec{k})+\omega_H(\vec{k})\right.\nonumber \\
&+&\left.\omega_G(\vec{k})
-4\omega_F(\vec{k})\right).
\end{eqnarray}
Performing the integrations in $V_{L}$ and subtracting the 
infinities give
\begin{eqnarray}
V_{L}(\rho_0)&=&{1\over 64\pi^2}(m_H^4\ln{m_H^2\over \kappa^2} 
+3m_Z^4\ln{m_Z^2\over \kappa^2}\nonumber \\
&+&m_G^4\ln{m_G^2\over \kappa^2}
-4m_t^4\ln{m_t^2\over \kappa^2})\nonumber \\
&-&{1\over 128\pi^2}(3m_H^4+5m_Z^4+3m_G^4-12m_t^4).
\end{eqnarray}
We remark that $V_{L}$ differs from $V_{VD}$ except
at the point of extremum where $\rho_0^2=2\mu^2$. 
At this
point, one has 
$m_G^2=0$ and $m_{\pm}^2
=m_Z^2$ which leads to $V_{VD}(\rho_0=2\mu^2)=V_L(\rho_0^2=2\mu^2)$. 
That 
$V_{VD}=V_L$ at the point of extremum is a consequence of Nielsen
identities\cite{niel}. 

To derive the Higgs mass bound from $V_{VD}(\rho_0)$ or $V_L(\rho_0)$,
one encounters a breakdown of the perturbation theory
at, for instance, 
${\lambda\over 16\pi^2}\ln{\lambda\rho_0^2\over \kappa^2}>1$
for a large $\rho_0$. To extend the validity of the effective potential for a
large $\rho_0$, the effective potential has to be improved by the 
renomalization group(RG) analysis. Let us denote the effective potential 
generically as $V_{eff}$. The renormalization-scale independence of $V_{eff}$
implies the following equation\cite{cw,BLW}:
\begin{eqnarray}
& &\left(
-\mu(\gamma_{\mu}+1){\partial \over \partial \mu}
+\beta_{\hat{g}}{\partial
\over \partial \hat{g}}\right. \nonumber \\
& &-\left. (\gamma_{\rho}+1)t{\partial \over \partial t}
+4\right)V_{eff}
(t\rho_0^i,\mu,\hat{g},\kappa)=0.
\end{eqnarray} 
where $\hat{g}$ denotes collectively the coupling constants $\lambda$, $g_1$,
$g_2$ and $y$; $\rho_0^i$ is an arbitrarily chosen initial value for $\rho_0$.
Solving this differential equation gives 
\begin{eqnarray}
& &V_{eff}(t\rho_0^i,\mu_i,\hat{g}_i,\kappa)=\exp\left(\int_0^{\ln t}
{4\over 1+\gamma_{\rho}(x)}dx\right)\nonumber \\
&\times & V_{eff}(\rho_0^i,\mu(t,\mu_i),
\hat{g}(t,\hat{g}_i),\kappa),
\label{IMPROVE}
\end{eqnarray}  
where
\begin{equation}
t{d\hat{g}\over dt}={\beta_{\hat{g}}(\hat{g}(t))\over 1+\gamma_{\rho}
(\hat{g}(t))} \ {\rm with} \ \hat{g}(0)=\hat{g}_i,
\label{BEGA}
\end{equation} 
and
\begin{equation}
\mu(t,\mu_i)=\mu_i\exp\left(-\int_0^{\ln t}
{1+\gamma_{\mu}(x)\over 1+\gamma_{\rho}(x)}dx\right)
\end{equation}
To fully determine $V_{eff}$ at large $\rho_0$, we need to calculate 
$\beta$ functions of $\lambda$, $g_1$, $g_2$ and $y$, and the anomalous
dimensions $\gamma_{\mu}$ and $\gamma_{\rho}$. It has been demonstrated 
that the $n$-loop effective potential is 
improved by $(n+1)$-loop $\beta$ and $\gamma$ functions\cite{ka,bkmn}. Since  
the effectve potential is calculated to the one-loop order, 
a consistent RG analysis
requires the knowledge of $\beta$ and $\gamma$ functions up to two loops.
As the computation of two-loop $\beta$ and $\gamma$ functions 
are quite involved, we will simply improve the tree-level 
effective potential with one-loop $\beta$ and $\gamma$ functions.   

We have the following one-loop $\beta$ and $\gamma$ functions in the
Landau-DeWitt gauge(we will set $\hbar=1$ from this point on):
\begin{eqnarray}
\beta_{\lambda}&=&{1\over 16\pi^2}\left({3\over 8}g_1^4-3\lambda g_1^2
-2y^4+4\lambda y^2+20\lambda^2\right),\nonumber \\
\beta_{g_1}&=&{g_1\over 4\pi^2}\left({g_1^2\over 16}-{g_1g_2\over 18}
+{g_2^2\over 27}\right),\nonumber \\
\beta_{g_2}&=&{g_2\over 4\pi^2}\left({g_1^2\over 16}-{g_1g_2\over 18}
+{g_2^2\over 27}\right),\nonumber \\
\beta_y&=&{y\over 8\pi^2}\left(y^2-{3g_1^2\over 8}+{g_1g_2\over 12}\right),
\nonumber \\  
\gamma_{\mu}&=&{1\over 2\pi^2}\left({3\lambda\over 4}+{3g_1^4\over 128}
-{3g_1^2\over 32}-{y^4\over 8\lambda}+{y^2\over 8}\right),\nonumber \\
\gamma_{\rho}&=&{1\over 64\pi^2}\left(-5g_1^2+4y^2\right).
\label{BETA}
\end{eqnarray}
We stress that
all the above functions except $\gamma_{\rho}$, the anomalous 
dimension of the scalar field, are in fact 
gauge-independent in the $\overline{MS}$ subtraction
scheme.
For $\gamma_{\rho}$ in the Landau gauge, we have  
\begin{equation}
\gamma_{\rho}=
{1\over 64\pi^2}\left(-3g_1^2+4y^2\right).
\end{equation}

\section{The Higgs Mass Bound}
The lower bound of the Higgs mass can be derived from the vacuum instability
condition for the effective potential. 
To derive the mass bound, one begins with Eq. (\ref{IMPROVE})
which implies
\begin{equation}
V_{tree}(t\rho_0^i,\mu_i,\lambda_i)={1\over 4}
\chi(t)\lambda(t,\lambda_i)
\left((\rho_0^i)^2-
2\mu^2(t,\mu_i)\right)^2,
\end{equation}
with $\chi(t)=
\exp\left(\int_0^{\ln t}
{4\over 1+\gamma_{\rho_0}(x)}dx\right)$. Since $\mu(t,\mu_i)$ decreases
as $t$ increases, we have $V_{tree}(t\rho_0^i,\mu_i,\lambda_i)
\approx {1\over 4}\chi(t)\lambda(t,\lambda_i)(\rho_0^i)^4$ for a sufficiently 
large $t$. Similarly, the one-loop effective potential 
$V_{1-loop}(t\rho_0^i,\mu_i,\hat{g}_i,\kappa)$ is also proportional to 
$V_{1-loop}(\rho_0^i,\mu(t,\mu_i),\hat{g}(t,\hat{g}_i),\kappa)$ with
the same proportional constant $\chi(t)$ as in $V_{tree}$. Since 
we shall neglect all running effects in $V_{1-loop}$, we have 
$\hat{g}(t,\hat{g}_i)=\hat{g}_i$ and $\mu(t,\mu_i)={1\over t}\mu_i$
in $V_{1-loop}$. For
a sufficiently large $t$, we can again approximate $V_{1-loop}$ by its quartic
terms. In the Landau-DeWitt gauge with the choice $\rho_0^i=\kappa$,
we have 
\begin{eqnarray}
V_{VD}\approx{(\rho_0^i)^4\over 64\pi^2}&[&9\lambda_i^2\ln(3\lambda_i)
+{g_{1i}^4\over 16}\ln({g_{1i}^2\over 4})-y_i^4\ln({y_i^2\over 2})\nonumber\\
&+&A_+^2(g_{1i},\lambda_i)\ln A_+(g_{1i},\lambda_i)\nonumber\\
&+&A_-^2(g_{1i},\lambda_i)\ln A_-(g_{1i},\lambda_i)\nonumber \\
&-&{3\over 2}(10\lambda_i^2+\lambda_i g_{1i}^2+{5\over 48}g_{1i}^4-y_i^4)],
\end{eqnarray}  
where $A_{\pm}(g_1,\lambda)=g_1^2/4+\lambda/2\cdot
(1\pm \sqrt{1+g_1^2/\lambda})$.
Similarly, the effective potential in the Landau gauge reads:
\begin{eqnarray}
V_{L}&\approx&{(\rho_0^i)^4\over 64\pi^2}[9\lambda_i^2\ln(3\lambda_i)
+{3g_{1i}^4\over 16}\ln({g_{1i}^2\over 4})-y_i^4\ln({y_i^2\over 2})
\nonumber \\
&+&\lambda_i^2\ln(\lambda_i)
-{3\over 2}(10\lambda_i^2+\lambda_i g_{1i}^2+{5\over 48}
g_{1i}^4-y_i^4)],
\end{eqnarray}  
Combining the tree level and the one-loop effective potential, we arrive at
\begin{equation}
V_{eff}(t\rho_0^i,\mu_i,\hat{g}_i,\kappa)\approx{1\over 4}\chi(t)
\left(\lambda(t,\lambda_i)+\Delta \lambda(\hat{g}_i)\right)(\rho_0^i)^4,
\end{equation}
where $\Delta \lambda$ denotes one-loop corrections given by Eqs. (30)
or (31). Let $t_{VI}=\rho_{VI}/\rho_0^i$. The condition for vacuum
instability of the effective potential is then\cite{ceq} 
\begin{equation}
\lambda(t_{VI},\lambda_i)+\Delta \lambda(\hat{g}_i)=0.
\label{VI}
\end{equation}  
We note that couplings $\hat{g}_i$ in $\Delta \lambda$ is evaluated 
at $\kappa=\rho_0^i$, which can be taken as the electroweak scale. Hence
$g_{1i}\equiv g/\cos\theta_W=0.67$, $g_{2i}\equiv 2e\tan\theta_W=0.31$,
and $y_i=1$. The running coupling $\lambda(t_{VI},\lambda_i)$ also depends on
$g_1$, $g_2$ and $y$ through $\beta_{\lambda}$, and $\gamma_{\rho}$ 
shown in Eq. (\ref{BETA}).

The strategy for solving Eq. (\ref{VI}) is to make an initial guess 
on $\lambda_i$, which enters into $\lambda(t)$ and $\Delta \lambda$, and
repeatedly adjusting $\lambda_i$ till $\lambda(t)$ completely 
cancells $\Delta \lambda$. 
For $t_{VI}=10^2$(or $\rho_0\approx 10^4$ GeV) which is the new-physics
scale reachable by LHC, we find $\lambda_i=4.83\times 10^{-2}$ 
for Landau-DeWitt gauge,
and $\lambda_i=4.8\times 10^{-2}$ for Landau gauge. For a higher 
instability scale such as the scale of grand unification, 
we have $t_{VI}=10^{13}$
or $\rho_0\approx 10^{15}$ GeV. In this case, we find $\lambda_i=3.13\times
10^{-1}$ for both Landau-DeWitt and Landau gauges. The numerical agreement 
between $\lambda_i$'s of two gauges can be attributed to
an identical $\beta$ function for the running of $\lambda(t)$, and
a small difference in $\Delta \lambda$ between two gauges.
We recall from Eq. (\ref{BEGA}) that the evolution of $\lambda$ in two
gauges will be different if effects of next-to-leading logarithm are
taken into account. In that case, the difference in $\gamma_{\rho}$ between 
two gauges give rise to different evolutions for $\lambda$. One may expect
to see non-negligible differences in $\lambda_i$ between two gauges for
a large $t_{VI}$. 

The critical value $\lambda_i=4.83\times 10^{-2}$ corresponds to a lower 
bound for the $\overline{MS}$ mass of the Higgs boson. Since $m_H=2\sqrt{
\lambda}\mu$, we have $(m_H)_{\overline{MS}}\geq 77$ GeV. For 
$\lambda_i=3.13\times 10^{-1}$, we have $(m_H)_{\overline{MS}}\geq 196$ GeV.
To obtain the lower bound for the physical mass of the Higgs boson,
finite radiative corrections must be added to the above bounds\cite{LW}. 
We will not pursue any further on these finite corrections since we are
simply dealing with a toy model. However we like to point out 
that this finite correction is gauge-independent as ensured by  
Nielsen identities\cite{niel}.   

\section{Conclusion}
We have computed the one-loop effective potential of 
an abelian $U(1)\times U(1)$
model in the Landau-DeWitt gauge, which reproduces the result
given by the gauge-independent 
Vilkovisky-DeWitt formalism. One-loop $\beta$ and $\gamma$ functions 
are also computed to facilitate the RG improvement of the effective 
potential. A gauge-independent lower bound for the Higgs self-coupling
or equivalently the $\overline{MS}$ mass of the Higgs boson is derived.
We compare this bound to that obtained by the ordinary effective potential
computed in Landau gauge. The numerical values of both bounds are
almost identical due to the leading-logarithmic approximation we have taken.
A complete next-to-leading analysis as well as an extension of this 
work to the full standard model will be reported in future publications.
  
\section*{Acknowledgements}
We thank W.-F. Kao for discussions.
This work is supported in part by
National Science Council of R.O.C. under grant numbers 
NSC 87-2112-M-009-038, and NSC 88-2112-M-009-002.


\section*{References}
\begin{thebibliography}{99}
%
\bibitem{jackiw}
R. Jackiw, \Journal{\PRD}{9}{1686}{1974}.
%
\bibitem{DJ}
L. Dolan and R. Jackiw, \Journal{\PRD}{9}{2904}{1974}.
%
\bibitem{niel}
N. K. Nielsen, \Journal{\NPB}{101}{173}{1975}.
%
\bibitem{BLW}
D. Boyanovsky, W. Loinaz and R. S. Willey, \Journal{\PRD}{57}{100}{1998}.
%
\bibitem{BBHL}
D. Boyanovsky, D. Brahm, R. Holman and D.-S. Lee, \Journal{\PRD}
{54}{1763}{1996}.
%
\bibitem{lc}
G.-L. Lin and T.-K. Chyi, NCTU-HEP-9804, hep-ph/9811213.
%
\bibitem{vil}
G. Vilkovisky, ``The Gospel According to DeWitt'' in {\em
Quantum Theory of Gravity}, ed. S. M. Christensen (Adam Hilger,
Bristol, 1983); \Journal{\NPB}{234}{215}{1984}.
%
\bibitem{dw2}
See, for example,  B. S.  DeWitt in 
{\em Quantum Field Theory and Quantum Statistics:
Essays in Honour of the 60th Birthday of E. S. Fradkin}, eds. 
I. A. Batalin, C. J. Isham and G. A. Vilkovisky (Hilger, Bristol, 1987), 
p. 191. 
%
\bibitem{com1}
A mere replacement of $-\eta^i$ with $\sigma^i(\Phi,\phi)$, as suggested
in Ref.\cite{vil}, is not satisfactory for calculations beyond one-loop, 
because $\Gamma[\Phi]$ does not generate one-particle irreducible
diagrams at the higher loops. A modified construction was given by DeWitt 
in Ref.\cite{dw2}. Since we will be only concerned with one-loop 
corrections, we shall adhere to the current construction 
in our subsequent discussions.    
%
\bibitem{kun}
See, for example,
G. Kunstatter, in {\em Super Field Theories}, eds. H. C. Lee, V. Elias,
G. Kunstatter, R. B. Mann, and K. S. Viswanathan(Plenum, New York, 1987), 
p. 503.
%
\bibitem{LW}
W. Loinaz and R. S. Willey, \Journal{\PRD}{56}{7416}{1997}. 
%
\bibitem{rebhan}
A. Rehban, \Journal{\NPB}{288}{832}{1987}; \Journal{\NPB}{298} 
{726}{1988}.
%
\bibitem{LSY} 
G.-L. Lin, H. Steger and Y.-P. Yao, \Journal{\PRD}{44}{2139}{1991}.
%
\bibitem{FT}
E. S. Fradkin and A. A. Tseytlin, \Journal{\NPB}{234}{509}{1984}.
%
\bibitem{rt}
I. H. Russell and D. J. Toms, \Journal{\PRD}{39}{1735}{1989}.
%
\bibitem{com2}
Although properties of this gauge are discussed in Ref.\cite{FT}, 
authors of this paper used other 
gauge to calculate the effective potential of scalar QED.
%
\bibitem{grassmann}
In the current gauge, $\tilde{D}^{-1}_{ij}=D^{-1}_{ij}$ 
since $\Gamma^i_{jk}$ vanishes. Furthermore,
for ghost fields, it is the ghost propagator rather than its inverse 
that will appear in
the effective action. This is due to the Grassmannian nature 
of ghost fields.
\bibitem{cw}
S. Coleman and E. Weinberg, \Journal{\PRD}{7}{1888}{1973}.
%
\bibitem{ka}
B. Kastening, \Journal{\PLB}{ 283}{287}{1992}.
%
\bibitem{bkmn} 
M. Bando, T. Kugo, 
N. Maekawa and H. Nakano, \Journal{\PLB}{301}
{83}{1993}.
%
\bibitem{ceq}
J. A. Casas, J. R. Espinosa and M. Quiros, 
\Journal{\PLB}{342}{171}{1995}.

\end{thebibliography}
\end{document}